\documentclass[preprint,showpacs,preprintnumbers,amsmath,amssymb]{revtex4}

\usepackage{graphicx}% Include figure files
\usepackage{dcolumn}% Align table columns on decimal point
\usepackage{bm}% bold math
\usepackage{epsfig}

\begin{document}

\preprint{APS/123-QED}

\title{Exact Site Percolation Thresholds Using the Site-to-Bond and Star-Triangle Transformations}% Force line breaks with \\

\author{Chris Scullard}
\email{scullard@uchicago.edu}
\affiliation{%
Department of Geophysical Sciences, University of Chicago, Chicago, Illinois 60637
}%

\date{July 26, 2005}% It is always \today, today,

\begin{abstract}
I construct a two-dimensional lattice on which the inhomogeneous site percolation threshold is exactly calculable and use this result to find two more lattices on which the site thresholds can be determined. The primary lattice studied here, the ``martini lattice'', is a hexagonal lattice with every second site transformed into a triangle. The site threshold of this lattice is found to be $0.764826...$, while the others have $0.618034...$ and $1/\sqrt{2}$. This last solution suggests a possible approach to establishing the bound for the hexagonal site threshold, $p_c<1/\sqrt{2}$. To derive these results, I solve a correlated bond problem on the hexagonal lattice by use of the star-triangle transformation and then, by a particular choice of correlations, solve the site problem on the martini lattice.
\end{abstract}

\pacs{Valid PACS appear here}% PACS, the Physics and Astronomy
\maketitle

\section{Introduction}
\label{intro}
The star-triangle transformation has been a useful tool in 2-d percolation theory \cite{Grimmett, Stauffer}. It was originally used by Sykes and Essam \cite{Sykes}, along with planar duality, to find the thresholds for inhomogeneous bond percolation on the triangular and hexagonal lattices, and was later adapted by Wierman \cite{Wierman} to solve the bond problem on the bowtie lattice and its dual. In fact, all exact results in 2-d are either derived directly from duality or matching properties, or rely on the star-triangle transformation in some way. In the present work, this method will be extended to problems with some limited correlation structure.

\begin{figure}
\begin{center}
\includegraphics[width=2.5in]{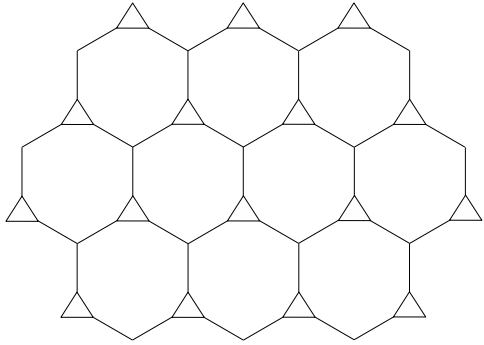}
\caption{The martini lattice, drawn to emphasize its origin as a hexagonal lattice with every second site transformed into a triangle.} \label{fig:martinilat2}
\end{center}
\end{figure}

\begin{figure}
\begin{center}
\includegraphics[width=2.5in]{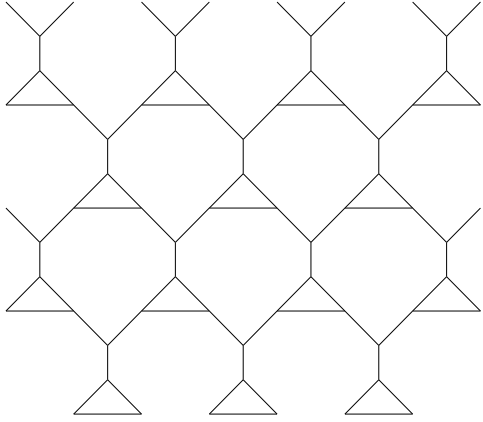}
\caption{The martini lattice, drawn to emphasize the martinis.} \label{fig:martinilat}
\end{center}
\end{figure}

The primary lattice studied here, which I call the ``martini lattice'' due to the shape of the basic cell, is the one shown in figures \ref{fig:martinilat2} and \ref{fig:martinilat}. Each site has 3 nearest neighbors, but the lattice is non-uniform because some sites are $(3,9^2)$ while others are $(9^3)$ in the notation of Gr\"{u}nbaum and Shephard \cite{Grunbaum}. This lattice is mentioned on page 186 of that book as an example of what they call a 2-homeohedral tiling of valence 3.

\begin{figure}[t]
\begin{center}
\includegraphics[width=2in]{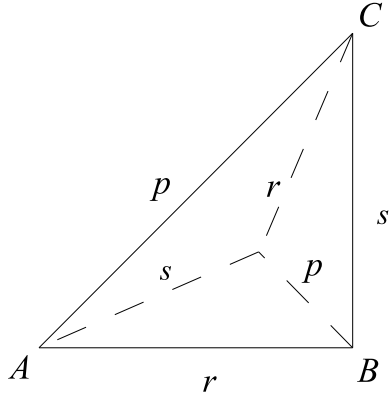}
\caption{The star-triangle transformation. $p,r,s$ denote the probabilities of their corresponding bonds on the triangle and star, and $A,B,C$ label the sites.} \label{fig:sttrans}
\end{center}
\end{figure}
\section{Star-Triangle Transformation}
\label{sec:2}
\begin{figure}[tbp]
\begin{center}
\includegraphics{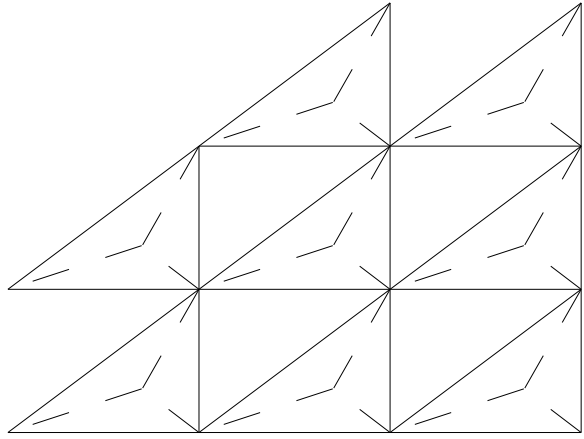}
\caption{The star-triangle transformation. Replacing each triangle with a dashed star transforms the triangular lattice into the hexagonal lattice.} \label{fig:trihex}
\end{center}
\end{figure}

The star-triangle transformation exploits the fact that if the bonds of a unit cell of the triangular lattice ({\bf T}) are replaced by a corresponding star, as illustrated in figure \ref{fig:sttrans}, the result is the hexagonal lattice ({\bf H}), which is the triangular lattice's dual (figure \ref{fig:trihex}). The bond percolation threshold of a lattice, $L$, and its dual, $L^d$, are related, in 2-d, by the well-known formula \cite{Grimmett}
\begin{equation}
p_c^{\mathrm{bond}}(L)=1-p_c^{\mathrm{bond}}(L^d)
\end{equation}
This means that the appearance of the infinite open cluster on $L$ coincides with the disappearance of the infinite closed cluster on $L^d$.
The star-triangle transformation leads to another relationship between the critical probabilities of $\bf{H}$ and $\bf{T}$ besides $p_c(\mathrm{\bf T})=1-p_c(\mathrm{\bf H})$, which allows both to be determined exactly. The method even works for the inhomogeneous case, where the probabilities of each bond being open on the base triangle are different, resulting in a critical surface rather than a critical point.

The argument proceeds as follows. Consider bond percolation on the triangle and superimposed star shown in figure \ref{fig:sttrans}. The probabilities $p, s, r$ refer to the probabilities that their corresponding bonds are open on either the star or triangle. We can ask several questions about the connectedness of the sites $A, B, C$. For example, what is the probability that $A$ is connected to both $B$ and $C$, an event we will denote $P(A \rightarrow B, A \rightarrow C)$, through open bonds on the triangle? This is easily found to be
\begin{equation}
P(A \rightarrow B, A \rightarrow C)=ps+pr(1-s)+sr(1-p)
\end{equation}
Next, we want the probability that $A, B, C$ are connected through {\it closed} bonds on the star. We denote this event $Q^*(A \rightarrow B, A \rightarrow C)$. $Q$ will hereafter denote the probability of events that happen in closed bonds and $*$ will indicate that the event happens by traversing the star rather than the triangle. Since this event only happens if all three bonds are closed, we have
\begin{equation}
Q^*(A \rightarrow B, A \rightarrow C)=(1-p)(1-r)(1-s)
\end{equation}
If we now consider the entire lattice, we can see that the replacement of triangles by stars turns the triangular lattice into the hexagonal lattice, i.e. the dual (figure \ref{fig:trihex}). This implies that the condition
\begin{equation}
P(A \rightarrow B, A \rightarrow C)=Q^*(A \rightarrow B, A \rightarrow C) \label{eq:critloc}
\end{equation}
defines our critical surface. The connectivity of open bonds on {\bf T} is exactly the same as that of the closed bonds on {\bf H} when this condition is satisfied. This means that if there is an infinite open cluster on {\bf T}, then there is an infinite closed cluster on {\bf H}. This leads, by our previous discussion of duality, to the conclusion that there is neither an infinite open cluster on {\bf T} nor an infinite closed cluster on {\bf H}, i.e. we are on the critical surface.
Substituting our results into (\ref{eq:critloc}) and simplifying, we are led to the final result:
\begin{equation}
psr-p-s-r+1=0 \label{eq:uncortriloc}
\end{equation}
Of course, setting $s=r=p$ leads to the critical point for bond percolation on the triangular lattice, $p_c=2 \sin \pi/18$. We can also check that the same result is obtained by studying different connectivities of the sites. For example, the probability that $A$ connects $B$, but not $C$, denoted $P(A \rightarrow B, A \nrightarrow C)$, is given by:
\begin{equation}
P(A \rightarrow B, A \nrightarrow C)=p(1-p)^2
\end{equation}
Also,
\begin{equation}
Q^*(A \rightarrow B, A \nrightarrow C)=p(1-p)^2
\end{equation}
so that $P(A \rightarrow B, A \nrightarrow C)=Q^*(A \rightarrow B, A \nrightarrow C)$ for all $p$, so this does not provide any constraint. If we calculate $P(A \nrightarrow B, A \nrightarrow C)=Q^*(A \nrightarrow B, A \nrightarrow C)$, we recover (\ref{eq:uncortriloc}) as we should.
\section{Correlated Bond Percolation on the Triangular Lattice}
\begin{figure}[tbp]
\begin{center}
\includegraphics[width=2in]{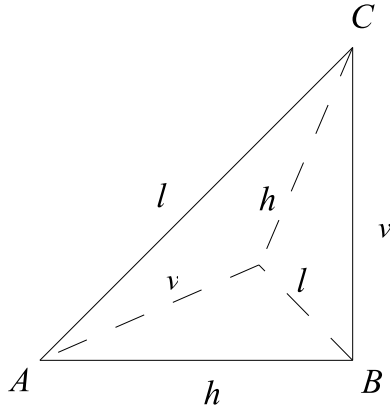}
\caption{Labels used in treating the correlated triangle. $A,B,C$ label sites and $v,h,l$ label bonds on the triangle and the corresponding bonds on the star - note that these are now names not probabilities.} \label{fig:cortri}
\end{center}
\end{figure}

The method of Sykes and Essam can be extended to the case where the bonds of the triangle are not independent. Our critical surface will now appear as a constraint between the $1$, $2$, and $3$-point joint probabilities. It is important to note that although all bonds in a triangle are correlated, there are no correlations between neighboring triangles, so a given bond is only correlated to two of its neighbors. The dual lattice is constructed in the same way as in the uncorrelated case, with each bond in the dual inheriting the probabilities and correlations of the original lattice. Labelling the bonds $v,h,l$ as shown in figure \ref{fig:cortri}, we will deal with the quantities $P(h,v,l)$, $P(h,l)$, $P(v,h)$, $P(v,l)$, $P(v)$, $P(h)$, $P(l)$, which are the set of $1$, $2$, and $3$-point joint probabilities of the indicated bonds being open. Probabilities of bonds being closed will be denoted with a bar over the bond name, e.g. $P(\bar{v})$. We can now repeat the procedure outlined above but with our joint probabilities:
\begin{equation}
P(A \rightarrow B, A \rightarrow C)=P(v,h)+P(v,l,\bar{h})+P(h,l,\bar{v})
\end{equation}
and
\begin{equation}
Q^*(A \rightarrow B, A \rightarrow C)=P(\bar{v},\bar{l},\bar{h})
\end{equation}
Equating these gives our critical surface:
\begin{equation}
P(v,h)+P(v,l,\bar{h})+P(h,l,\bar{v})-P(\bar{v},\bar{l},\bar{h})=0 \label{eq:cortriloc0}
\end{equation}
There are many equivalent ways this can be expressed. For example, if we use the condition 
\begin{equation}
P(A \nrightarrow B, A \nrightarrow C)=Q^*(A \nrightarrow B, A \nrightarrow C) \nonumber
\end{equation}
we obtain the more compact
\begin{equation}
P(v)+P(\bar{v},h,l)-P(\bar{h},\bar{l})=0 \label{eq:cortriloc}
\end{equation}
Although they look dissimilar, equations (\ref{eq:cortriloc0}) and (\ref{eq:cortriloc}) are in fact the same constraint and it is a simple matter to relate them using identities.

To compare with our earlier results, it is easy to see that setting $P(\bar{h},\bar{l})=(1-r)(1-s)$, $P(\bar{v},h,l)=(1-p)rs$, $P(v)=p$ leads to (\ref{eq:uncortriloc}).

\section{Site-to-Bond Transformation}
\label{sec:1}
\begin{figure}
\begin{center}
\includegraphics[width=2.5in]{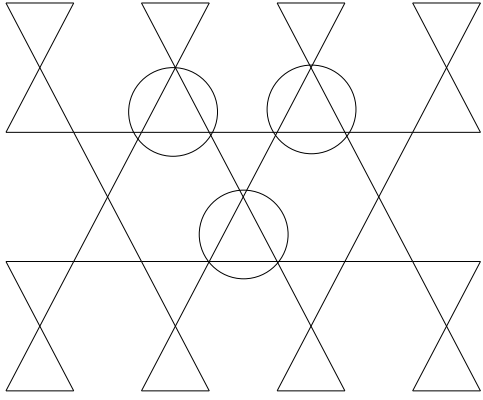}
\caption{The Kagom\'{e} lattice. The circled triangles are the ones on which we apply the site-to-bond transformation.}\label{fig:kagome}
\end{center}
\end{figure}

If we consider a realization of site percolation on a given lattice, we can transform it into a bond process by declaring a bond to be open if both its bounding sites are occupied, and closed otherwise. By doing this we introduce correlations between neighboring bonds; the probability that a given bond is open is $p^2$, but the probability that a bond is open given that one of its neighbors is open is $p$. Furthermore, there are also three-point correlations; the probability that a bond is open is $1$ if two of its neighbors on opposite ends of the bond are open. It is clear that the existence or lack of an infinite open cluster are properties that will be shared by both the site and transformed bond problems.
\begin{figure}
\begin{center}
\includegraphics[width=1.5in]{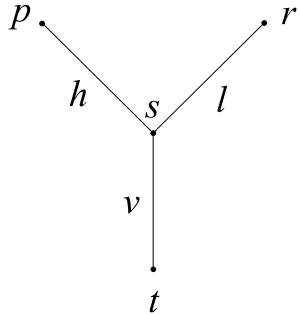}
\caption{The assignment of probabilities to the sites on the martini lattice.} \label{fig:martinistar}
\end{center}
\end{figure}
If we now consider sites on a triangle, we can use these rules to derive joint probabilities for the bonds, which we can then use in the criticality condition (\ref{eq:cortriloc}). It is easy to see that if the sites are occupied with probability $p$:
\begin{eqnarray}
P(v)&=&p^2 \label{eq:kag1}\\
P(\bar{h},\bar{l})&=&(1-p)^3+3p(1-p)^2+p^2(1-p) \label{eq:kag2}\\
P(h,l,\bar{v})&=&0 \label{eq:kag3}
\end{eqnarray}
If we use these in (\ref{eq:cortriloc}), we do not expect to have solved the site problem on the triangular lattice. The critical surface is not appropriate to that problem because we have not included correlations between triangles. I suggest that the threshold we will discover is that of the site problem on the Kagom\'{e} lattice (figure \ref{fig:kagome}). To see this, consider the triangles outlined in figure \ref{fig:kagome}. Clearly they are not correlated to each other if we use the site-to-bond transformation. But percolation of bonds on these triangles implies percolation of the lattice since, due to $3$-point correlations, the bonds on the separating triangles will be open with probability $1$ if two bonds on either side of it are open. Plugging expressions (\ref{eq:kag1})-(\ref{eq:kag3}) into (\ref{eq:cortriloc}) we obtain the polynomial
\begin{equation}
1-3p_c^2+p_c^3=0 \label{eq:kagomecrit}
\end{equation}
with solution $p_c=1-2 \sin \pi/18$ which is indeed the critical threshold of the Kagom\'{e} lattice. However, since this is just the covering lattice of {\bf H}, its threshold has long been known through more elementary means.

We can obtain our new results by considering percolation on the star. The critical surface in this case will be given by the complement of (\ref{eq:cortriloc}):
\begin{equation}
P(\bar{v})+P(v,\bar{h},\bar{l})-P(h,l)=0 \label{eq:corstarloc}
\end{equation}
In fact, we will consider the inhomogeneous site problem, and assign probabilities $p$,$r$,$s$,$t$, to the sites as shown in figure \ref{fig:martinistar}.
\begin{eqnarray}
P(\bar{v})&=&1-st\\
P(h,l)&=&prs\\
P(v,\bar{h},\bar{l})&=&st(1-p)(1-r)
\end{eqnarray}
This leads to the critical surface
\begin{equation}
1-rst-prs-pst+stpr=0 \label{eq:martiniloc}
\end{equation}
which is the central result of this work. But to what lattice does it correspond? In the previous example, where we obtained the Kagom\'{e} lattice, we needed to insert extra triangles to separate our correlated triangles. Inserting these separating triangles in between the stars, we obtain the martini lattice shown in figure \ref{fig:martinilat}. The critical threshold for site percolation is obtained by setting $r=s=t=p$:
\begin{equation}
1-3p_c^3+p_c^4=0
\end{equation}
which has solution on $[0,1]$:
\begin{equation}
p_c=0.764826...
\end{equation}
We can obtain further results by making different choices for the probabilities. For example, by setting $s=1$, $r=t=p$, we expect the Kagom\'{e} lattice to reappear, since $s=1$ turns the star back into a triangle. Plugging these into (\ref{eq:martiniloc}) we indeed get (\ref{eq:kagomecrit}). Other choices are possible that lead to a variety of results.

\begin{figure}[htbp]
\begin{center}
\includegraphics[width=2.5in]{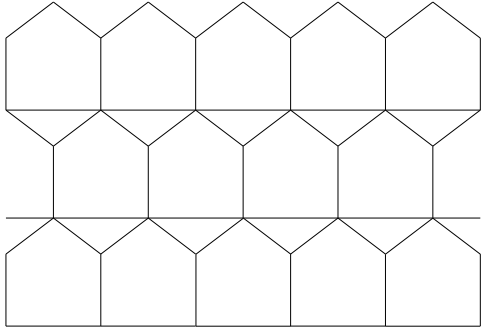}
\caption{The lattice obtained from the martini lattice by setting $t=r=1$. On this lattice, $p_c^{\mathrm{site}}=(\sqrt{5}-1)/2=0.618034...$ and $p_c^{\mathrm{bond}}=1/2$.} \label{fig:houses}
\end{center}
\end{figure}

\subsection{$t=r=1$}

The corresponding lattice is the one shown in figure \ref{fig:houses} and it resembles a stack of houses or a neighborhood. It is non-uniform, with some sites $(3^2,5^3)$ with $5$ nearest neighbors and others $(3,5^2)$ with $3$ nearest neighbors, and falls somewhere between the hexagonal lattice and the $(3^3,4^2)$ lattice - almost exactly between as we will see. The $(3^2,5^3)$ sites have probability $s$ and the $(3,5^2)$ sites $p$. The critical locus is
\begin{equation}
1-s-ps=0
\end{equation}
Setting $s=p$ leads to the critical threshold
\begin{equation}
1-p_c-p_c^2=0
\end{equation}
or $p_c=(\sqrt{5}-1)/2=0.618034...$

This lattice has some interesting properties that are worth mentioning. For one, it is self-dual, meaning we can immediately locate its bond threshold at $1/2$, as well as the site threshold of its covering lattice. Also, it seems to be an intermediate step in transforming the hexagonal lattice into the $(3^3,4^2)$ lattice. If we remove the horizontal bonds, the ``floors'' of the houses, we get the hexagonal lattice. If instead we symmetrically add another horizontal bond, a ``crossbeam'' under the roof, we get the $(3^3,4^2)$ lattice. It has been found numerically \cite{Suding} that $p_c(3^3,4^2)\approx 0.550$ and $p_c(\mathrm{\bf H}) \approx .697$ and their average is around $0.624$, which is similar to the result for the present lattice. Apparently, adding the floor has roughly the same effect on the threshold as adding the crossbeam.

\subsection{$s=t=1$}
Setting $s=t=1$ and $r=p$, we get the covering lattice of the square bond problem, leading to $p_c=1/2$, which is a very roundabout way of solving that problem.

\subsection{$r=1$}
\begin{figure}
\begin{center}
\includegraphics[width=2.5in]{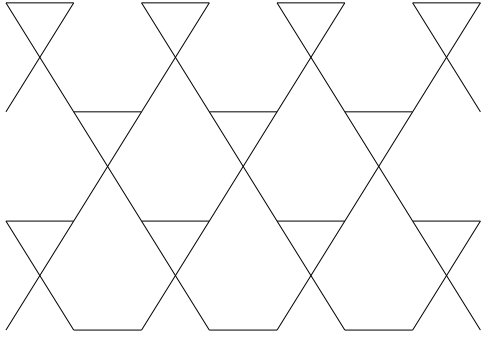}
\caption{The lattice obtained by setting $r=1$. The site threshold is $p_c=1/\sqrt{2}$.} \label{fig:nearhex}
\end{center}
\end{figure}

Again, a star is turned into a triangle, but a different one from that which produced the Kagom\'{e} lattice earlier. The lattice that results here is shown in figure \ref{fig:nearhex}. The critical surface is
\begin{equation}
1-st-ps=0
\end{equation}
Setting $s=t=p$ yields
\begin{equation}
1-2p_c^2=0
\end{equation}
which means $p_c=1/\sqrt{2}=0.707...$. This is an interesting result for several reasons. For one, some sites have $3$ nearest neighbors while others have $4$. Thus, if we were to make a guess strictly on the basis of nearest neighbors, we might be led to believe that this lattice's site threshold was smaller than that of the hexagonal lattice, where every site has $3$ nearest neighbors. The contrary is true however, as we know from numerical results \cite{Suding}. Interestingly, $p_c=1/\sqrt{2}$ was once conjectured to be the exact site threshold for the hexagonal lattice \cite{Kondor} but was shortly thereafter judged unlikely from numerical considerations \cite{Djordjevic}.
\begin{figure}
\begin{center}
\includegraphics{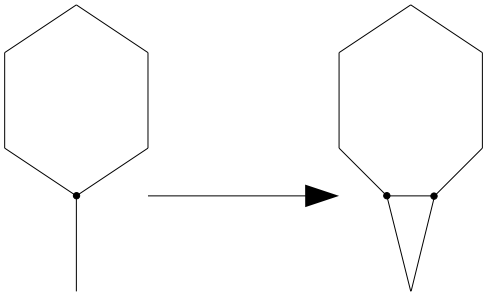}
\caption{The transformation that takes the hexagonal lattice to the one shown in figure \ref{fig:nearhex}. The site at the bottom of every hexagon is divided into two sites, and a bond is inserted between them.}\label{fig:hextrans}
\end{center}
\end{figure}

This result suggests an approach to finding an upper bound for $p_c^{\mathrm{site}}({\bf H})$. Consider the following procedure for producing this lattice. Starting with the usual hexagonal lattice, take the site at the bottom of each hexagon, split it into two sites and connect them together with a bond, as shown in figure \ref{fig:hextrans}. All that is required is to show that this procedure always increases the critical probability and we will have shown that
\begin{equation}
p_c^{\mathrm{site}}({\bf H})<\frac{1}{\sqrt{2}}
\end{equation}
Of course, even better would be if the shift in the critical point could be exactly determined. It remains to be seen whether either of these suggestions is actually feasible.

\section{Concluding Remarks}
We have found three lattices whose site percolation thresholds can be calculated exactly, and shown that one of these solutions might lead to an improved upper bound for the hexagonal lattice.

It should be noted that a special case of  (\ref{eq:martiniloc}) has previously appeared in the literature, though in a slightly different context. In considering a mixed site/bond problem, Kondor \cite{Kondor} derived an expression that matches (\ref{eq:martiniloc}) for $r=t=p$. The situation he considered was bond percolation on the triangular lattice in which there is an ``activating'' site in the middle of every second triangle that turns all its surrounding bonds off if it is unoccupied. This problem is evidently isomorphic to site percolation on the martini lattice. 

Also, the value $1/\sqrt{2}$ for the lattice in figure \ref{fig:nearhex} is actually somewhat puzzling. Usually, simple values like this indicate an elementary tranformation from a known lattice with $p_c=1/2$. If such a transformation exists in this case, it does not appear to be obvious. 

\begin{acknowledgments}
I would like to thank Robert Ziff for many helpful comments and for providing some encouraging numerical results.
I am also deeply indebted to Bruce Buffett, without whom this work would not have been possible.
\end{acknowledgments}

%\newpage %Just because of unusual number of tables stacked at end
\bibliography{scullardPRE}% Produces the bibliography via BibTeX.

\begin{thebibliography}{8}
\expandafter\ifx\csname natexlab\endcsname\relax\def\natexlab#1{#1}\fi
\expandafter\ifx\csname bibnamefont\endcsname\relax
  \def\bibnamefont#1{#1}\fi
\expandafter\ifx\csname bibfnamefont\endcsname\relax
  \def\bibfnamefont#1{#1}\fi
\expandafter\ifx\csname citenamefont\endcsname\relax
  \def\citenamefont#1{#1}\fi
\expandafter\ifx\csname url\endcsname\relax
  \def\url#1{\texttt{#1}}\fi
\expandafter\ifx\csname urlprefix\endcsname\relax\def\urlprefix{URL }\fi
\providecommand{\bibinfo}[2]{#2}
\providecommand{\eprint}[2][]{\url{#2}}

\bibitem[{\citenamefont{Grimmett}(1991)}]{Grimmett}
\bibinfo{author}{\bibfnamefont{G.}~\bibnamefont{Grimmett}},
  \emph{\bibinfo{title}{Percolation}} (\bibinfo{publisher}{Taylor and Francis,
  London}, \bibinfo{year}{1991}).

\bibitem[{\citenamefont{Stauffer}(1982)}]{Stauffer}
\bibinfo{author}{\bibfnamefont{D.}~\bibnamefont{Stauffer}},
  \emph{\bibinfo{title}{Introduction to Percolation Theory}}
  (\bibinfo{publisher}{Springer-Verlag, Berlin}, \bibinfo{year}{1982}).

\bibitem[{\citenamefont{Sykes and Essam}(1964)}]{Sykes}
\bibinfo{author}{\bibfnamefont{M.~F.} \bibnamefont{Sykes}} \bibnamefont{and}
  \bibinfo{author}{\bibfnamefont{J.~W.} \bibnamefont{Essam}},
  \bibinfo{journal}{J. Math. Phys.} \textbf{\bibinfo{volume}{5}},
  \bibinfo{pages}{1117} (\bibinfo{year}{1964}).

\bibitem[{\citenamefont{Wierman}(1984)}]{Wierman}
\bibinfo{author}{\bibfnamefont{J.~C.} \bibnamefont{Wierman}},
  \bibinfo{journal}{J. Phys. A: Math. Gen.} \textbf{\bibinfo{volume}{17}},
  \bibinfo{pages}{1525} (\bibinfo{year}{1984}).

\bibitem[{\citenamefont{Gr\"{u}nbaum and Shephard}(1987)}]{Grunbaum}
\bibinfo{author}{\bibfnamefont{B.}~\bibnamefont{Gr\"{u}nbaum}}
  \bibnamefont{and} \bibinfo{author}{\bibfnamefont{G.~C.}
  \bibnamefont{Shephard}}, \emph{\bibinfo{title}{Tilings and Patterns}}
  (\bibinfo{publisher}{Freeman, New York}, \bibinfo{year}{1987}).

\bibitem[{\citenamefont{Suding and Ziff}(1999)}]{Suding}
\bibinfo{author}{\bibfnamefont{P.~N.} \bibnamefont{Suding}} \bibnamefont{and}
  \bibinfo{author}{\bibfnamefont{R.~M.} \bibnamefont{Ziff}},
  \bibinfo{journal}{Phys. Rev. E} \textbf{\bibinfo{volume}{60}},
  \bibinfo{pages}{275} (\bibinfo{year}{1999}).

\bibitem[{\citenamefont{Kondor}(1980)}]{Kondor}
\bibinfo{author}{\bibfnamefont{I.}~\bibnamefont{Kondor}}, \bibinfo{journal}{J.
  Phys. C: Solid St. Phys.} \textbf{\bibinfo{volume}{13}},
  \bibinfo{pages}{L531} (\bibinfo{year}{1980}).

\bibitem[{\citenamefont{Djordjevic et~al.}(1982)\citenamefont{Djordjevic,
  Stanley, and Margolina}}]{Djordjevic}
\bibinfo{author}{\bibfnamefont{Z.~V.} \bibnamefont{Djordjevic}},
  \bibinfo{author}{\bibfnamefont{E.~H.} \bibnamefont{Stanley}},
  \bibnamefont{and}
  \bibinfo{author}{\bibfnamefont{A.}~\bibnamefont{Margolina}},
  \bibinfo{journal}{J. Phys. A: Math. Gen.} \textbf{\bibinfo{volume}{15}},
  \bibinfo{pages}{L405} (\bibinfo{year}{1982}).

\end{thebibliography}

\end{document}